\documentclass[apj]{emulateapj}

\usepackage{graphicx,times}
\usepackage{subfigure}
\newcommand{\be}{\begin{equation}}
\usepackage{threeparttable}
\usepackage{booktabs}
\newcommand{\ee}{\end{equation}}
\newcommand{\bea}{\begin{eqnarray}}
\newcommand{\eea}{\end{eqnarray}}

\usepackage{enumerate}
\usepackage{amsmath}
\usepackage{cases}
\usepackage{longtable}
\usepackage{hyperref}
\usepackage{epstopdf}
\usepackage{amsmath,bm}
\usepackage{amssymb}
\usepackage{natbib}
\usepackage{morefloats}
\usepackage{multirow}
\usepackage{array}
\usepackage{verbatim}
\usepackage{setspace}

\begin{document}

\title{Adiabatic non-resonant acceleration in magnetic turbulence and hard spectra of gamma-ray bursts}

\author{Siyao Xu\altaffilmark{1} and Bing Zhang \altaffilmark{1,2,3}}

\altaffiltext{1}{Department of Astronomy, School of Physics, Peking University, Beijing 100871, China; syxu@pku.edu.cn}
\altaffiltext{2}{Kavli Institute for Astronomy and Astrophysics, Peking University, Beijing 100871, China}
\altaffiltext{3}{Department of Physics and Astronomy, University of Nevada Las Vegas, NV 89154, USA; zhang@physics.unlv.edu}

\begin{abstract}

We introduce a non-resonant
acceleration mechanism arising from the second adiabatic invariant 
in magnetic turbulence and apply it to study the prompt emission spectra of gamma-ray bursts (GRBs).  
The mechanism contains both the first- and second-order Fermi acceleration, 
originating from the interacting turbulent reconnection and dynamo processes.
It leads to a hard electron energy distribution up to a cutoff energy at the balance between the acceleration and synchrotron cooling. 
The sufficient acceleration rate ensures a rapid hardening of any initial energy distribution to a power-law distribution with the index $p \sim 1$, 
which naturally produces a low-energy photon index $\alpha \sim -1$ via the synchrotron radiation. 
For typical GRB parameters, the synchrotron emission can extend to a characteristic photon energy on the order of $\sim 100$ keV.  
 
\end{abstract}


\section{Introduction}

The gamma-ray burst (GRB) prompt emission is closely related to the physics of particle acceleration and radiation. 
The origin of its spectral behavior, despite the empirical description \citep{Band93}, 
has not been well understood. 
The observed low-energy photon index has a typical value $\alpha \sim -1$ \citep{Pre00,Ka06,Zhang11,Nava11}, 
which is difficult to reconcile with the standard model invoking the first-order Fermi acceleration and
fast synchrotron cooling \citep{Pre02,Ghi00,KZ15}.
Many attempts have been made to seek the solution to the problem
(e.g., \citealt{Bra94, Li97, Me00, PeZ06, Asano09, Asano11, Daigne11, UZ14}).

In either a Poynting-flux-dominated or a baryonic relativistic outflow,
turbulence is inevitably present and participates in the electron acceleration process. 
The stochastic acceleration through resonant scattering with magnetic fluctuations has been used to explain the hard electron spectrum 
(e.g. \citealt{Byk96, Asano09,Asano11,Murase12}).
Advances in turbulence theories 
\citep{GS95,LV99}
provide new insight into the problem. 
Turbulent reconnection,
which was put forward by 
\citet{LV99}
and numerically confirmed in both 
non-relativistic 
\citep{KowL09,KL12}
and relativistic 
\citep{Tak15}
plasmas, provides an efficient dissipation mechanism of the magnetic energy in the GRB outflow.
\cite{Zh11} invoked a moderately Poynting-flux-dominated GRB jet and collision-induced magnetic dissipation
to interpret GRB prompt emission. This ICMART model envisages
significant turbulent reconnection and reconnection-driven turbulence in the emission region of GRBs.
Relativistic MHD simulations \citep{Deng15} and Monte Carlo simulations 
\citep{ZZ14} confirmed some features (e.g. efficient energy dissipation, existence of mini-jets and their
effects on the lightcurves) of the original model \citep{Zh11}. 
The large emission radius invoked in the ICMART model allows a modification
of the fast synchrotron cooling theory through invoking the decrease of the magnetic field in the emission region
as the jet expands in space, which can reproduce the desired $\alpha \sim -1$ even for first-order-Fermi-accelerated
electrons (probably through reconnection)
\citep{UZ14}.

In magnetohydrodynamic (MHD) turbulence, the turbulent reconnection efficiently relaxes tangled field lines and facilitates turbulent 
motions. Meanwhile, turbulent shearing motions stretch field lines and generate magnetic fluctuations via the turbulent dynamo
\citep{XL16}. 
Their nonlinear interactions regulate the dynamics of MHD turbulence 
and affect the acceleration of 
the electrons for which the second adiabatic invariant applies
\citep{BruL16}. 
The adiabatic condition is easily satisfied in a strongly magnetized GRB outflow, 
because either the gyroresonance scattering is absent with 
the particle Larmor radius below turbulence scales, 
or it is inefficient due to turbulence anisotropy 
\citep{YL02}. 
It is the first-order Fermi process within each reconnection/dynamo region
and the second-order Fermi process as particles stochastically encounter the reconnection/dynamo event. 
The stochastic nature originates from the balance between the annihilation 
and generation of magnetic fluxes in a trans-Alfv\'{e}nic turbulence
\citep{GS95} (hereafter GS95).
In this Letter, based on the modern understanding of the dynamical nature of MHD turbulence, 
we analytically solve the evolution of the electron energy distribution resulting from 
the above adiabatic acceleration in trans-Alfv\'{e}nic turbulence
(\S 2), 
and demonstrate that the resultant hard energy distribution entails a hard synchrotron spectrum at low energies 
in the prompt GRB phase,
consistent with observations (\S 3). 
A discussion of the results is in \S 4.

\section{Adiabatic acceleration of electrons in MHD turbulence}

\subsection{Energy spectrum of electrons}

We consider a turbulence regime with the magnetic and kinetic energies in equipartition. 
It is the trans-Alfv\'{e}nic turbulence described by 
GS95,
and has been numerically tested in both non-relativistic/low-$\sigma$ 
\citep{MG01,CLV_incomp}
and relativistic/high-$\sigma$ 
\citep{Cho05,Cho14}
cases.
In trans-Alfv\'{e}nic turbulence, 
there co-exist the magnetic field line-stretching process, i.e., turbulent dynamo
\citep{CVB09,XL16},
driven by turbulent velocities and the field line-shrinking process driven by the turbulent magnetic reconnection
\citep{LV99}. 
These two opposing processes take place at the same rate over all the turbulent scales, with the 
overall magnetic flux conserved.

The electrons, when they are not subject to scattering by magnetic fluctuations,
undergo the first-order Fermi acceleration in reconnection regions 
\citep{DeG05,LaO09}
and deceleration in dynamo regions as a consequence of the second adiabatic invariant,
leading to a globally diffusive energy gain.
It is similar to the process that moving particles are stochastically trapped between approaching ``mirrors" 
and receding ``mirrors"
\citep{Flue}.

The energy gain/loss within each turbulent eddy 
follows the first-order Fermi process. 
Despite the large energy change, we consider the Fokker-Planck equation as a valid description,
since it yields the basically identical particle spectrum as that from the statistical approach independent of the energy increment
\citep{Sch93}.

The evolution equation of the energy distribution function is
\begin{equation}\label{eq: gin}
\begin{aligned}
   \frac{\partial N}{\partial t} =&  a_2 \frac{\partial}{\partial E} \Big(E\frac{\partial (EN)}{\partial E}\Big)
   - (a_{1,\text{rec}}-a_{1,\text{dyn}}) \frac{\partial (EN)}{\partial E}  \\
  &  + \beta \frac{\partial (E^2N)}{\partial E},
\end{aligned}
\end{equation}
where $N(E,t)dE$ is the number of electrons within the energy interval from $E$ to $E+dE$. 
The terms on the RHS represent the second- and first-order Fermi processes, and synchrotron loss, 
while the adiabatic expansion of the plasma and electron escape are neglected.

We consider that the turbulent eddies at the 
injection scale $l_\text{tur}$ of the GS95 turbulence, i.e. the typical energy-containing scale, 
dominate the reconnection/dynamo. 
Provided $r_L < l_\text{tur}$, where $r_L$ is the Larmor radius, 
the stochastic acceleration rate $a_2$, 
related to the comparable reconnection acceleration rate $a_{1,\text{rec}}$ and the dynamo deceleration rate $a_{1,\text{dyn}}$, 
is independent of particle energy.
It is associated with the eddy turnover rate,  
\begin{equation}\label{eq: a2}
    a_2 \sim  a_{1,\text{rec}} \sim  a_{1,\text{dyn}} \sim \xi \frac{u_\text{tur}}{l_\text{tur}},
\end{equation}
where $u_\text{tur}$ is the relativistic turbulent velocity at $l_\text{tur}$, 
and $\xi = \Delta E / E \sim \gamma_\text{tur}^2$ 
for highly relativistic turbulence and particles to account for the energy conversion efficiency,
with the turbulence Lorentz factor $\gamma_\text{tur}$
\citep{Flue}.

Therefore, Eq. \eqref{eq: gin} can be reduced to 
\begin{equation}
   \frac{\partial N}{\partial t} =  a_2 \frac{\partial}{\partial E} \Big(E\frac{\partial (EN)}{\partial E}\Big)
    + \beta \frac{\partial (E^2N)}{\partial E}.
\end{equation}
After the substitution of the relations, 
$f = EN =  \exp(-\epsilon E) u(x, \tau)$, $\epsilon = \beta / a_2$, $x=\ln E$, $\tau = a_2 t$, and some algebra, we derive 
\begin{equation}\label{eq: simf}
     \frac{\partial u}{\partial \tau} = \frac{\partial^2 u}{\partial x^2} - \frac{E}{E_\text{cf}} \frac{\partial u}{\partial x},
\end{equation}
where we define the cutoff energy corresponding to the balance between the stochastic acceleration and the synchrotron loss,
\begin{equation}\label{eq: cutofe}
   E_\text{cf} = \frac{a_2}{\beta}  = \frac{3 (m_e c^2)^2 a_2}{4 \sigma_T c U_B}, 
\end{equation}
where $\sigma_T$ is the Thomson cross section, $c$ is the light speed, $m_e$ is the 
electron rest mass, and $U_B = B^2 / (8 \pi)$ is the magnetic energy density. 
Obviously in the energy range $E \ll E_\text{cf}$, Eq. \eqref{eq: simf} becomes a straightforward diffusion equation,
\begin{equation}\label{eq: funde}
     \frac{\partial u}{\partial \tau} = \frac{\partial^2 u}{\partial x^2} ,
\end{equation}
which allows us to analyze the time-dependent behavior of $N(E,\tau)$.

The general form of the solution to Eq. \eqref{eq: funde} is (e.g., \citealt{Ev98}),
\begin{equation}\label{eq: gen}
     u(x, \tau) =  \frac{1}{2 \sqrt{\pi \tau}}  \int_{y_l}^{y_u}   \exp\Big[-\frac{(x - y)^2}{4 \tau}\Big] u(y, 0) dy ,
\end{equation}
with the initial functional form $u(y,0)$ within the range [$y_l$, $y_u$].
The Gaussian function shows that 
the energy distribution spreads out in energy space following $E \sim \exp(\pm2\sqrt{\tau})$
\citep{Mel69}. 
Within a finite range of $E$ with lower and upper limits $E_l (=\exp(y_l) )$ and $E_u (=\exp(y_u))$, 
there is 
\begin{equation}\label{eq: tlu}
    E_u = E_l \exp (2 \sqrt{\tau_{lu}}), ~~ \tau_{lu} = \frac{(\ln E_u -\ln E_l)^2}{4}.
\end{equation}
After the time $\tau_{lu}$, 
the initial spectral form is essentially smeared out within the range, 
the behavior of $u$ is independent of $x$, and thus the energy spectrum of electrons
\begin{equation}\label{eq: hars}
      N(E,\tau) = E^{-1} u(\tau) \exp\Big(-\frac{E}{E_\text{cf}}\Big)
\end{equation}
has a universal form of $E^{-1}$ at $E<E_\text{cf}$.
The synchrotron cooling has a negligible effect on the energy distribution in the lower energy range away from $E_\text{cf}$.

\subsection{Examples for different initial energy distributions} 

(1) Delta function

Starting from an initial point source of energy with 
$u(y, 0) = \delta(y-x_0), y\in(-\infty, +\infty), x_0 = \ln E_0$, $u(x, \tau)$ evolves as   
\begin{equation}
    u(x, \tau) = \frac{1}{2 \sqrt{\pi \tau}} \exp\Big[-\frac{(x - x_0)^2}{4 \tau}\Big], 
\end{equation}
and thus 
\begin{equation}\label{eq: anad}
          N(E, \tau)  =  E^{-1} \frac{1}{2 \sqrt{\pi \tau}} \exp\Big[-\frac{(\ln E - \ln E_0)^2}{4 \tau}\Big]  \exp\Big(-\frac{E}{E_\text{cf}}\Big).
\end{equation}
The Gaussian component has a negligible contribution to the spectral form 
at a sufficiently large $\tau$.

(2) Power-law function

Given an initially steeper spectrum with the power-law index $p_0>1$, 
\begin{equation}\label{eq: inpw}
   N(E, 0) = C E^{-p_0} \exp \Big(-\frac{E}{E_\text{cf}}\Big), ~~ E \in (E_l, E_u), 
\end{equation}
that is, $u(y, 0) = C \exp [(1-p_0)y]$, where $C$ is an arbitrary constant, 
we can obtain the evolving spectrum by using Eq. \eqref{eq: gen}, 
\begin{equation}\label{eq: ipwn}
\begin{aligned}
    N(E, \tau) = & E^{-1}  \frac{C}{2 \sqrt{\pi \tau}}  \exp\Big(-\frac{E}{E_\text{cf}}\Big)  \\
 &   \int_{y_l}^{y_u}   \exp\Big[-\frac{(\ln E - y)^2}{4 \tau}\Big] \exp [(1-p_0)y] dy , 
\end{aligned}
\end{equation}
Its asymptotic form at a short time ($\tau \ll 1$) is 
\begin{equation}\label{eq: aset}
    N(E, \tau) = E^{-p_0} \frac{C}{2 \sqrt{\pi \tau}} \exp\Big(-\frac{E}{E_\text{cf}}\Big), 
\end{equation}
which is governed by the initial power-law shape at $E<E_\text{cf}$. 
Its long-time ($\tau \sim \tau_{lu}$) asymptotic expression is
\begin{equation}\label{eq: aslt}
    N(E, \tau) = E^{-1} \frac{C (E_u^{1-p_0} - E_l^{1-p_0})}{2 (1-p_0) \sqrt{\pi \tau}}  \exp\Big(-\frac{E}{E_\text{cf}}\Big) ,
\end{equation}
which again recovers the universal $E^{-1}$ power-law distribution at $E<E_\text{cf}$.

\begin{figure*}[htbp]
\centering
\subfigure[Initial delta function]{
   \includegraphics[width=8cm]{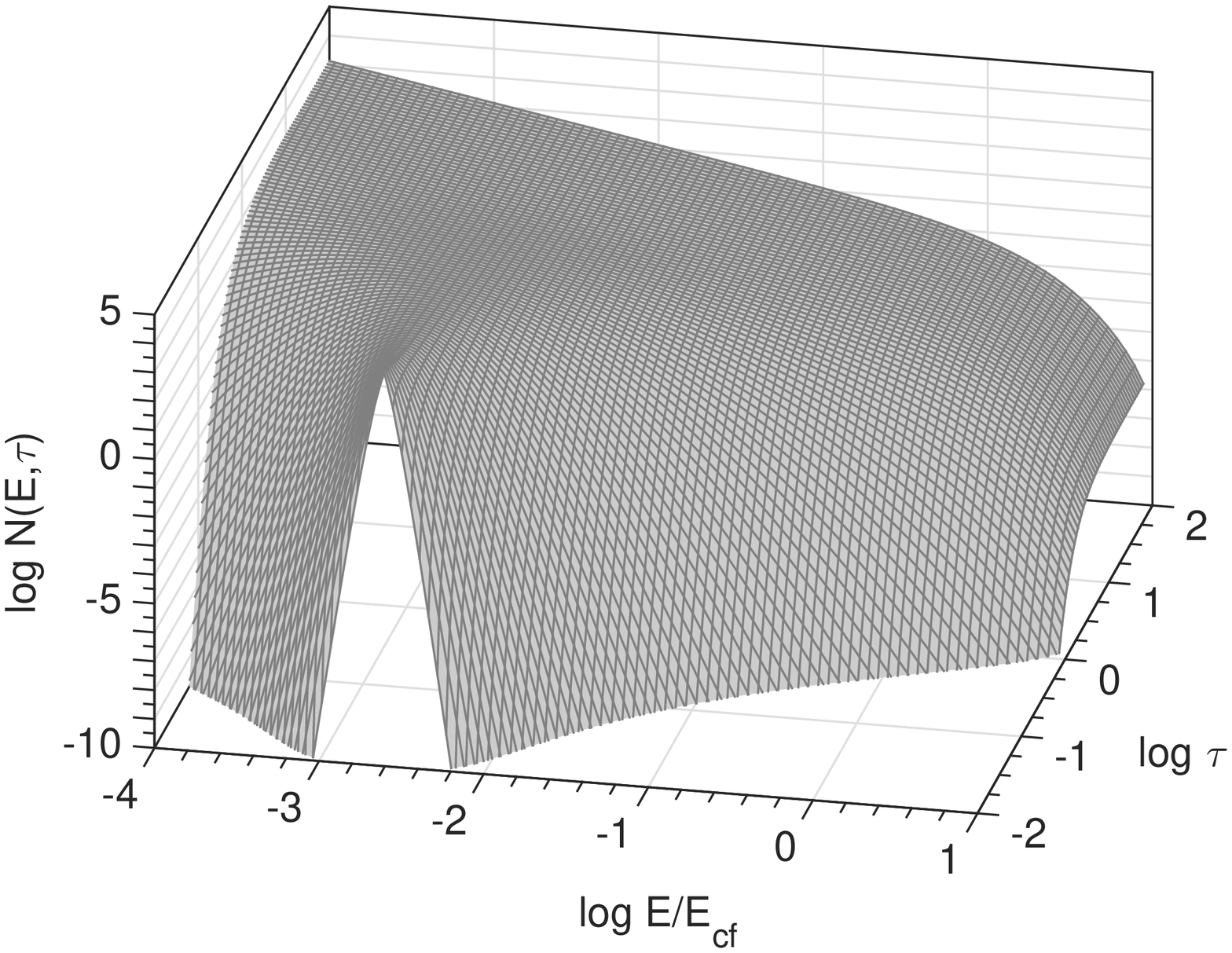}\label{fig: delt3}}
\subfigure[Initial power-law function]{
   \includegraphics[width=8cm]{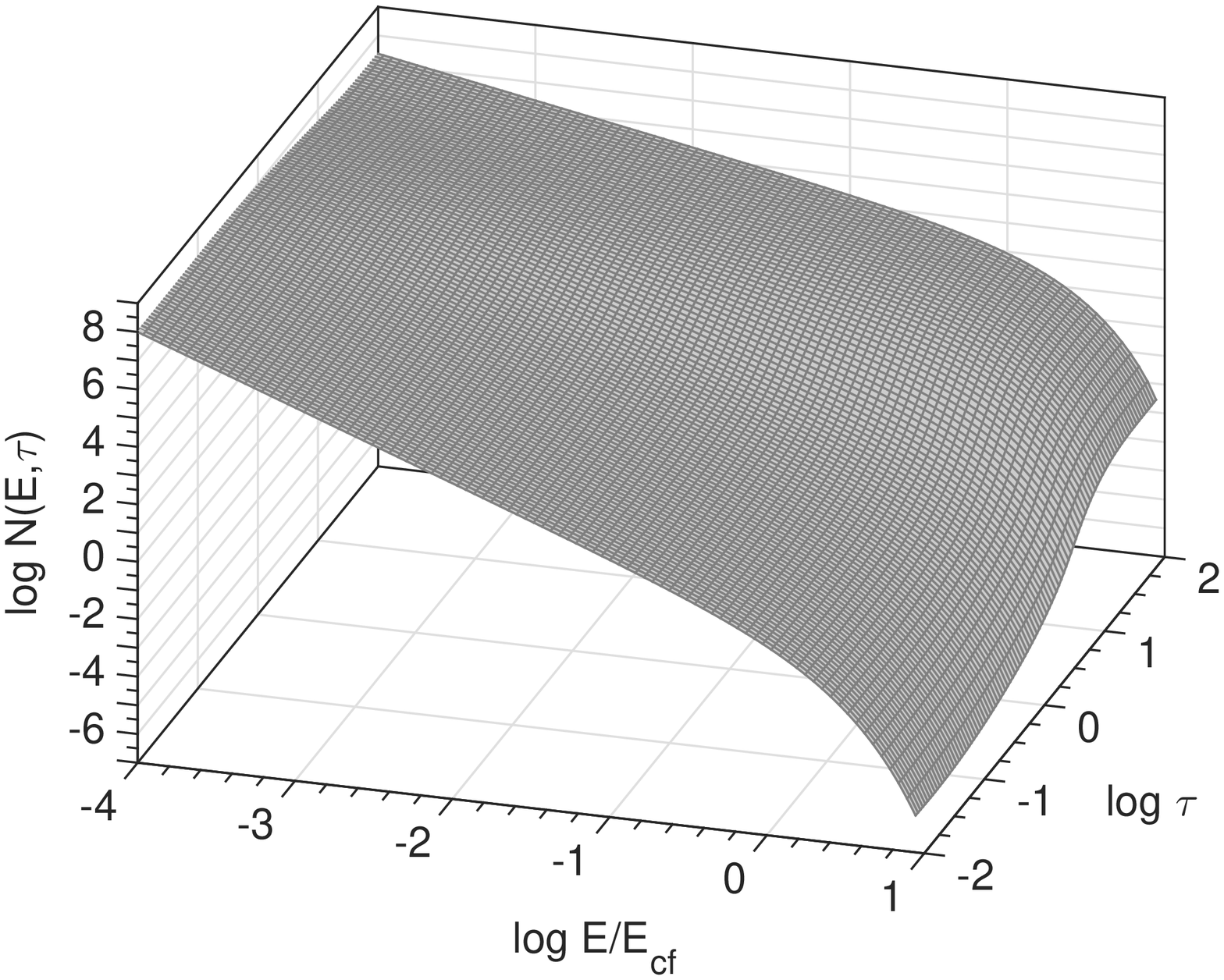}\label{fig: pow3}}  
\subfigure[Initial delta function]{
   \includegraphics[width=8cm]{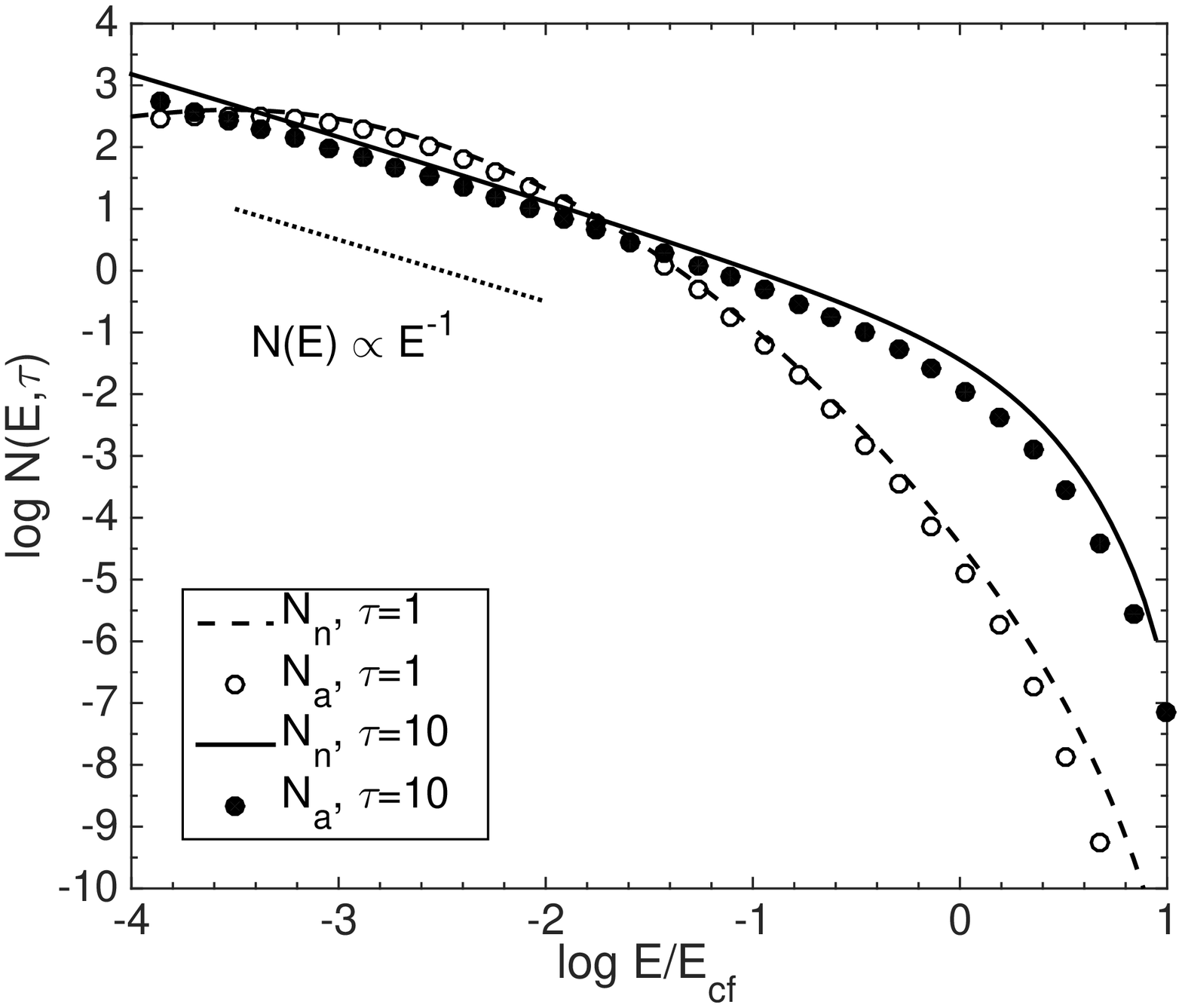}\label{fig: del2t}}
\subfigure[Initial power-law function]{
   \includegraphics[width=8cm]{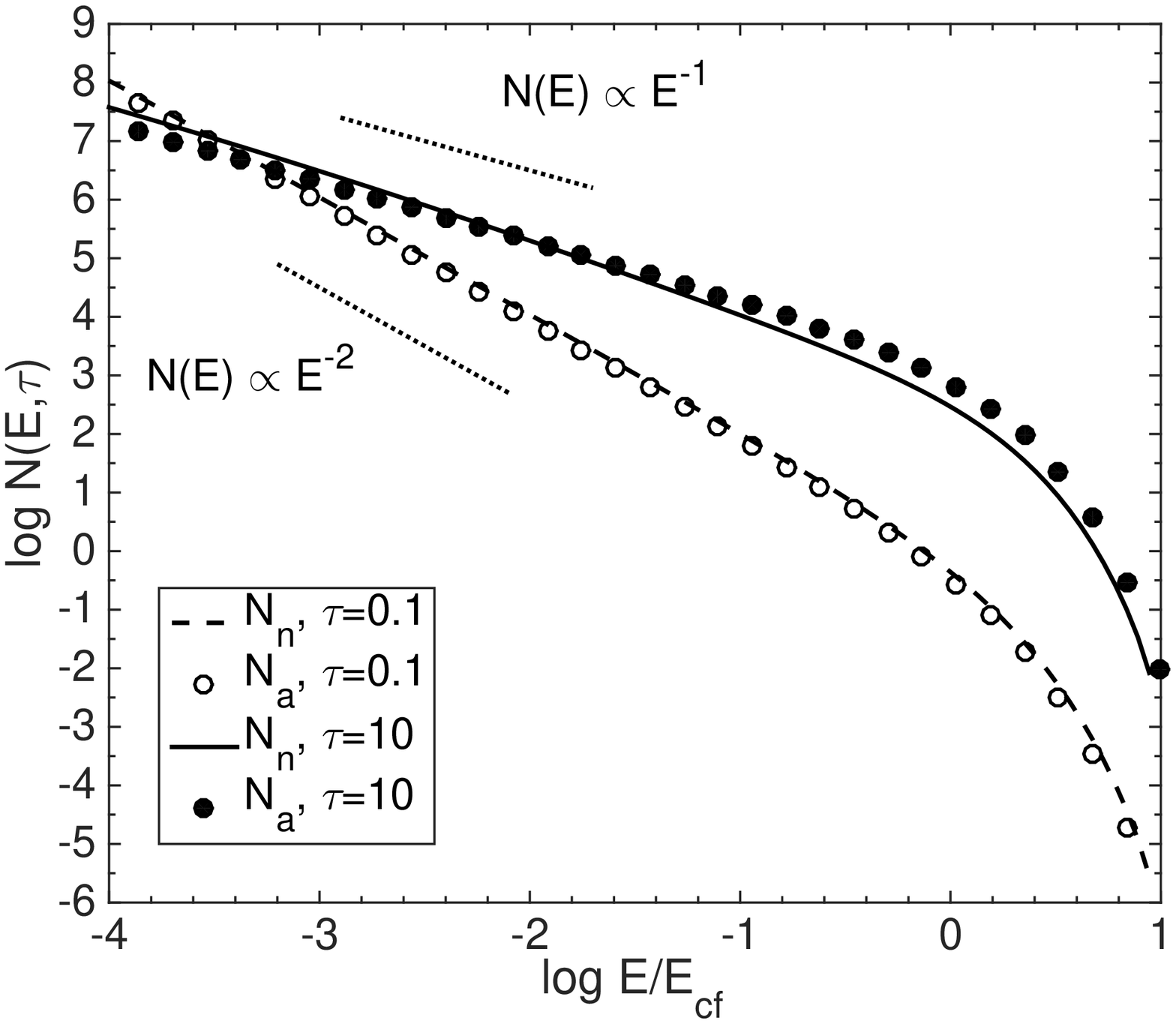}\label{fig: pw2t}}  
\caption{(a), (b) Temporal evolution of the energy distribution of electrons from the 
numerical solution to Eq. \eqref{eq: simf}. 
(c), (d) The spectral distribution at a specified $\tau$. ``$N_n$" and ``$N_a$" denote the numerical and analytical results, respectively.
The analytical results in (c) are from Eq. \eqref{eq: anad}, 
and in (d) are from Eq. \eqref{eq: aset} for $\tau=0.1$ and Eq. \eqref{eq: aslt} for $\tau=10$.}
\label{fig: 3map}
\end{figure*}

In Fig. \ref{fig: 3map}, we display the electron energy distributions 
obtained from the numerical solution to Eq. \eqref{eq: simf}. 
As illustrative examples, we adopt a delta function at $E_0/E_\text{cf} = 2.4 \times 10^{-3}$ in Fig. \ref{fig: delt3}, 
a power-law function with $C=1$ and $p_0 = 2$ over the entire energy range presented in Fig. \ref{fig: pow3}, and set $\epsilon=1$. 
Fig. \ref{fig: del2t} and \ref{fig: pw2t} show the asymptotic analytical solutions in the low-energy limit at a specified $\tau$, 
which agree well with the numerical results. 
As expected, irrespective of the initial spectral form, the distribution broadens and shifts in energy space and eventually 
conforms to the universal power-law shape $E^{-1}$ at $E<E_\text{cf}$ over the timescale of 
$\tau_{lu} \approx 20$ (Eq. \eqref{eq: tlu}).

\begin{figure}[htbp]
\centering
   \includegraphics[width=8cm]{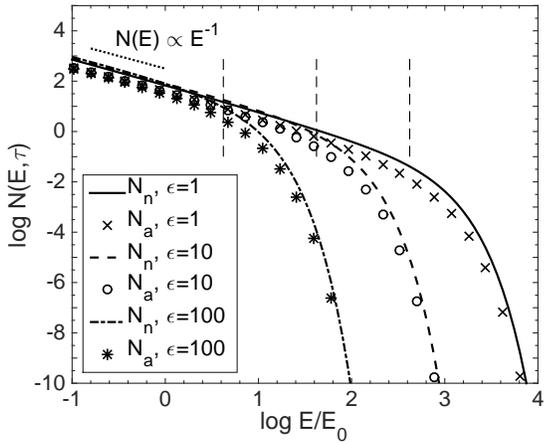}
\caption{The energy distribution with different $\epsilon$ values at $\tau = 10$. The vertical dashed lines indicate 
the positions of $E_\text{cf}$.}
\label{fig: del2e}
\end{figure}

To examine the effect of synchrotron cooling on the energy distribution, 
in Fig. \ref{fig: del2e} we present the results with an initial delta function
and different $\epsilon$ values at $\tau=10$. 
Notice that unlike in Fig. \ref{fig: 3map} where $E$ is normalized by $E_\text{cf}$, here we use the normalization $E/E_0$ to show 
the change of the cutoff energy with varying $\epsilon$. 
As $\epsilon$ increases, the spectral cutoff moves to a lower energy, 
but the distribution below the cutoff energy is unaffected and remains the $E^{-1}$ form. 

\section{Synchrotron emission of the accelerated electrons}

The above adiabatic acceleration of electrons arising in trans-Alfv\'{e}nic turbulence 
leads to a hard electron energy distribution (see Eq. \eqref{eq: hars}),
corresponding to 
\begin{equation}
     N(\gamma_e) \sim \gamma_e^{-p}  \exp\Big(-\frac{\gamma_e}{\gamma_{e,\text{cf}}}\Big), ~~ p =1, 
\end{equation}
where $p$ is the power-law index, $\gamma_e$ is the electron Lorentz factor, 
and $\gamma_{e,\text{cf}}  =  E_\text{cf} / m_e c^2$.

According to the relation between $p$ and the index $\alpha$ of the synchrotron photon number spectrum 
\citep{Ry79},
there is 
\begin{equation}
    N(\nu) \sim \nu^\alpha \exp \Big(-\Big(\frac{\nu}{\nu_c}\Big)^\frac{1}{2}\Big), ~~
   \alpha =  -\frac{p+1}{2}  = -1,
\end{equation}
where the $\delta$-function approximation for the single-electron spectrum is made. 
This is the typical low-energy photon spectral index of GRBs \citep{Pre00,Ka06,Zhang11,Nava11}.

Unlike a soft distribution with $p>2$, for which the lower cutoff energy accounts for the characteristic electron energy and 
synchrotron frequency
\citep[e.g.][]{Zh04,Pir04},
as regards a hard distribution, the upper cutoff energy $E_\text{cf}$ is more significant as it 
dominates the electron energy density,
and may characterize  
the peak energy of the $\nu F_\nu \propto \nu^2 N(\nu)$ spectrum
\citep{DaiC01}.
$E_\text{cf}$ depends on the acceleration mechanism (Eq. \eqref{eq: a2}, \eqref{eq: cutofe}), 
\begin{equation}\label{eq: ecfac}
   E_\text{cf} =  \frac{\xi u_\text{tur}}{l_\text{tur} \beta}  = \frac{6\pi \xi (m_e c^2)^2}{\sigma_T B^2 l_\text{tur}}, 
\end{equation}
where $u_\text{tur}$ is approximately equal to $c$ for relativistic turbulence.

In the case of reconnection-driven turbulence, the thickness of the turbulent region increases with time, as indicated by the numerical 
results in  
\citet{Kow17}.
Meanwhile, the magnetic field strength $B$ decays with time 
\citep{PeZ06,UZ14,ZhL14}.
The comoving-frame $B$ during the GRB prompt phase is estimated as
\citep[e.g.][]{Zhang02}
\begin{equation}\label{eq: zmag}
    B = \sqrt{\frac{2}{c}} L^\frac{1}{2} r^{-1} \Gamma^{-1}, 
\end{equation}
where $L$ is the total outflow luminosity of the GRB, 
$r$ is the distance of the emission region from the central engine,
and $\Gamma$ is the Lorentz factor of the outflow.
We assume the following relation between $l_\text{tur}$ and $B$ to 
account for their anticorrelation, 
\begin{equation}\label{eq: ltb}
   l_\text{tur} = l_\text{0} \Big(\frac{B}{B_0}\Big)^{-\zeta}, ~ \zeta>0,
\end{equation}
where $l_0$ and $B_0$ are normalization parameters. 
Then $E_\text{cf}$ in Eq. \eqref{eq: ecfac} can be expressed as 
\begin{equation}
    E_\text{cf} = \frac{6\pi \xi (m_e c^2)^2}{\sigma_T B_0^\zeta l_\text{0}} B^{\zeta-2}.
\end{equation}
The corresponding electron Lorentz factor is 
\begin{equation}
    \gamma_{e,\text{cf}} =  \frac{6\pi \xi m_e c^2}{\sigma_T B_0^\zeta l_\text{0}} B^{\zeta-2},
\end{equation}
and the emitted photon energy in the observer frame is 
\begin{equation}
\begin{aligned}
    E_\text{s,obs}  & = (h \nu_\text{cf})_\text{obs} \\
   &= \hbar \frac{e B}{m_e c} \gamma_{e,\text{cf}}^2 \Gamma (1+z)^{-1}    \\    
   & = \hbar m_e c^3 e  \bigg(\frac{6\pi \xi}{\sigma_T B_0^\zeta l_0}\bigg)^2  \Gamma (1+z)^{-1} B^{2\zeta -3},
\end{aligned}
\end{equation}
where $h$ is the Planck constant, $e$ is the electron charge, and $z$ is the redshift.
Inserting Eq. \eqref{eq: zmag}, the above equation becomes 
\begin{equation}\label{eq: origen}
\begin{aligned}
    E_\text{s,obs}  =&  2^{\zeta-\frac{3}{2}}\hbar m_e c^{-\zeta+\frac{9}{2}} e  \bigg(\frac{6\pi \xi}{\sigma_T B_0^\zeta l_0}\bigg)^2  \\
   &  \Gamma^{-2\zeta+4} (1+z)^{-1} L^{\zeta-\frac{3}{2}} r^{-2\zeta+3}.
\end{aligned}
\end{equation}

The dependence of $E_\text{s,obs}$ on $\Gamma$, $L$, and $r$ is determined by the exact value of $\zeta$.
Based on the empirical tight correlation among the peak energy, $\Gamma$, and $L$ suggested by observations 
\citep{Lia15}, 
we adopt $\zeta \simeq 2.1$. 
By together assuming 
$\xi =10^4$, $B_0 = 10^5$G, $l_0 = 2\times10^9$cm, and using other typical parameters $\Gamma=100 \Gamma_{2}$, 
$L=10^{52} {\rm erg~s^{-1}} L_{52}$, $r=10^{15} \text{cm} r_{15}$, Eq. \eqref{eq: origen} gives  
\begin{equation}
   E_\text{s,obs} \simeq 385 \text{keV}~ \Big(\frac{1+z}{2}\Big)^{-1} \Gamma_2^{-0.2} L_{52}^{0.6} r_{15}^{-1.2}.
\end{equation}

Besides, we can also estimate the timescale for an initial energy distribution with the index $p_0$ to evolve to a hard spectrum
(Eq. \eqref{eq: a2}, \eqref{eq: tlu}, \eqref{eq: zmag}, \eqref{eq: ltb}),
\begin{equation}
\begin{aligned}
    t_{lu} = \frac{\tau_{lu}}{a_2} &=  \frac{1}{4 \xi}  \Big[\ln \Big(\frac{E_u}{E_l}\Big)\Big]^2 \frac{l_0}{u_\text{tur}} \Big(\frac{B}{B_0}\Big)^{-\zeta}  \\
     & =  \frac{1}{4 \xi} \Big(\frac{2}{c}\Big)^{-\frac{\zeta}{2}}  \Big[\ln \Big(\frac{E_u}{E_l}\Big)\Big]^2 \frac{B_0^\zeta l_0}{u_\text{tur}} \Gamma^\zeta L^{-\frac{\zeta}{2}} r^\zeta   \\
     & =  3 \times10^{-2}\text{s}~F(E_u,E_l)_2 \Gamma_{2}^{2.1} L_{52}^{-1.05} r_{15}^{2.1} ,
\end{aligned}
\end{equation}   
with $[\ln (E_u/E_l)]^2 =10^2 F(E_u,E_l)_2$.
Irrespective of the value of $p_0$ (which can be larger or smaller than one), 
after $t_{lu}$, the electron distribution index $p$ approaches one 
under the effect of the adiabatic acceleration.

\section{Discussion}

We have applied the adiabatic acceleration mechanism in MHD turbulence, 
which has been earlier identified by \cite{BruL16},
and derived a robust electron energy distribution index $p \sim 1$ and a synchrotron low-energy photon index $\alpha \sim -1$, generally consistent with the observations.
The estimated characteristic synchrotron emission energy with proper turbulence parameters required 
is in the sub-MeV regime, which is also consistent with the observations. A hard particle spectrum due to stochastic accelerations was also discussed within the GRB context by, e.g. \cite{Byk96, Asano09,Asano11,Murase12}. 
However, here we consider a different non-resonant acceleration 
related to the reconnection and dynamo processes in MHD turbulence,
and strictly derive $p=1$ analytically.

Depending on the relation between the magnetic and turbulent kinetic energies, turbulence has various regimes. 
In the magnetic energy-dominated turbulence, 
the first-order Fermi acceleration during the turbulent reconnection 
\citep{DeG05, Kow12}
can dominate the electron acceleration and shape the initial energy distribution. 
With the conversion of magnetic energy to turbulent kinetic energy, the acceleration process becomes 
globally stochastic in the trans-Alfv\'{e}nic turbulence 
and rapidly flattens the electron energy distribution. 
To more realistically model the synchrotron spectrum, 
one should consider the interplay between the particle injection and acceleration, 
with synchrotron cooling incorporated self-consistently (S. Xu et al. 2017, in preparation).


Besides the GRB prompt emission spectrum, observations of  
active galactic nuclei, blazars, and pulsar wind nebulae also reveal a hard electron distribution 
(e.g., \citealt{She06,Ha15}).
The acceleration mechanism presented here is 
a promising candidate for interpreting the spectral hardness in various scenarios.

\acknowledgments

We thank the anonymous referee for insightful comments. 
SX is grateful for the support from the 
Pilot-B program for gravitational wave astrophysics of the Chinese Academy of Sciences and 
the Research Corporation for Scientific Advancement during her visit at the Aspen Center for Physics. 
SX thanks Yuanpei Yuan for valuable discussions.
This work is partially supported by the National Basic Research Program (973 Program) of China under grant No. 2014CB845800.

\end{document}